\documentclass[conference]{IEEEtran}
\IEEEoverridecommandlockouts
\usepackage{cite}
\usepackage{amsmath,amssymb,amsfonts}
\usepackage{algorithmic}
\usepackage{graphicx}
\usepackage{textcomp}
\usepackage{xcolor}
\usepackage{url}
\usepackage{hyperref}
\usepackage{placeins}
\usepackage[utf8]{inputenc}
\usepackage{amsmath}
\usepackage{amssymb}
\usepackage{upgreek} 
\def\BibTeX{{\rm B\kern-.05em{\sc i\kern-.025em b}\kern-.08em
    T\kern-.1667em\lower.7ex\hbox{E}\kern-.125emX}}
\begin{document}

\title{Modeling Behavioral Signals in Job Scams: A Human-Centered Security Study}

\author{
\IEEEauthorblockN{
Goni Anagha\textsuperscript{1},
Vishakha Dasi Agrawal\textsuperscript{1},
Gargi Sarkar\textsuperscript{*},
Kavita Vemuri,
Sandeep Kumar Shukla
}
\IEEEauthorblockA{
International Institute of Information Technology Hyderabad,  
Gachibowli, Hyderabad, India 500 032\\
Email: \{goni.anagha@students., vishakha.agrawal@students., gargi.sarkar@research., kvemuri@, sandeeps@\}iiit.ac.in
}
\thanks{\textsuperscript{1}Goni Anagha and Vishakha Agrawal contributed equally to this work.}
\thanks{\textsuperscript{*}Corresponding author: Gargi Sarkar (gargi.sarkar@research.iiit.ac.in)}
}

\maketitle

\begin{abstract}
Job scams have emerged as a rapidly growing form of cybercrime that manipulates human decision-making processes. Existing countermeasures primarily focus on scam typologies or post-loss indicators, offering limited support for early-stage intervention. In this study, we examine how behavioral decision signals can be operationalized as computational features for identifying vulnerability-associated signals in job fraud. Using anonymous survey data collected from a university population, we analyze two dominant job scam pathways: payment-based scams that require upfront fees and task-based scams that begin with small rewards before escalating to financial demands. Drawing on behavioral economics, we operationalize sunk cost influence, urgency/time-pressure cues, and social proof as measurable behavioral signals, and analyze their association with payment behavior using exact inference under sparsity and uncertainty-aware estimation, with social proof treated as a context-dependent legitimacy cue rather than a standalone predictor. Our results show that urgency/time-pressure cues are significantly associated with payment behavior, consistent with their role as proximal compliance triggers during escalation. In contrast, opportunity-loss/FOMO cues were not reliably identifiable under the current operationalization in our encounter subset, highlighting the importance of measurement fidelity and cue-definition consistency. We further observe that emotional tone in victim narratives and selective non-response to sensitive questions vary systematically with financial loss and reporting behavior, suggesting that missingness may reflect a combination of survey fatigue and selective non-disclosure for sensitive items rather than purely random noise. These findings suggest that incorporating human behavioral signals into cybercrime detection and warning systems can improve early-stage risk assessment without intrusive monitoring.
\end{abstract}

\begin{IEEEkeywords}
Job scams, Behavioral economics, Cybercrime, Sunk cost influence, Urgency cues, Fear of missing out (FOMO), Social proof, Cognitive biases
\end{IEEEkeywords}

\section{Introduction}

Cybercrime has emerged as one of the most pervasive security and economic challenges of the digital era, inflicting substantial harm on individuals, businesses, and critical infrastructure worldwide. Within this broader landscape, employment fraud, commonly referred to as job scams, has emerged as one of the most rapidly expanding and financially consequential categories of online deception. Unlike technically complex attacks such as malware or network intrusions, job scams primarily operate through persuasive interaction, exploiting human judgment, trust, and decision-making processes. As a result, they frequently bypass traditional detection mechanisms that focus on technical artifacts or post-loss indicators, positioning job scams as a critical human-centered security challenge.

\subsection{The Rising Threat of Job Fraud}

 According to the Internet Crime Complaint Center (IC3) of the Federal Bureau of Investigation, more than 859,000 cybercrime complaints were recorded in 2024 alone, with reported losses exceeding \$16.6 billion, an increase of 33\% compared to the previous year \cite{fbi2024}. Over the five-year period from 2020 to 2024, IC3 documented over 4.2 million complaints and \$50.5 billion in cumulative losses, with daily complaint volumes rising from roughly 2,000 per month to nearly 2,000 per day \cite{fbi2024}. These figures illustrate not only the scale but also the increasing sophistication of cyber-enabled crime.

Recent data from the U.S. Federal Trade Commission (FTC), one of the most comprehensive global sources for fraud reporting, shows that losses from job scams increased from \$90 million in 2020 to \$501 million in 2024, representing a more than five-fold rise, alongside a tripling in the number of reported cases \cite{ftc2024}. In just the first six months of 2024, reported losses exceeded \$220 million \cite{ftc2024}. Although these figures originate from the U.S. context, they are widely used as indicators of international cybercrime trends due to the FTC’s rigorous reporting mechanisms. Moreover, prior research consistently demonstrates that fraud victimization is significantly underreported worldwide, often due to shame, embarrassment, or fear of judgment \cite{button2009, titus1995}. Consequently, the true scale of job fraud is likely substantially larger than reported statistics suggest.

Contemporary job scams have adapted to digitally mediated platforms such as WhatsApp, Telegram, encrypted messaging applications, and fraudulent job portals, leveraging the anonymity and global reach of the internet to target individuals during periods of economic vulnerability \cite{sarkar2025}. Modern employment fraud encompasses several distinct typologies, including Pay-to-Get-a-Job scams that require upfront payments for training, certification, or equipment; Work-from-Home schemes promising unrealistic income with minimal effort; Multi-Level Marketing (MLM) disguised as employment; and, most alarmingly, Task-Based scams, a gamified variant that has driven much of the recent surge in reported losses.

Task-Based scams represent a particularly concerning trend, with reports increasing from zero in 2020 to 5,000 in 2023, and then quadrupling to approximately 20,000 in just the first half of 2024, accounting for 38.8\% of all job scam reports as stated in the FTC report \cite{ftc2024}. These scams typically begin with unsolicited messages offering vague online work opportunities involving activities such as ``product boosting,'' ``app optimization,'' or content rating. Victims initially receive small, real payments for completing simple tasks, creating a false sense of legitimacy. The scam subsequently escalates into an extraction phase, requiring victims to deposit their own money, often in cryptocurrency, to unlock higher-level tasks and promised earnings that never materialize. Cryptocurrency losses to job scams reached \$41 million in the first half of 2024, nearly double the \$21 million reported in all of 2023 \cite{ftc2024}. This progression highlights how scammers exploit both psychological vulnerabilities and emerging payment technologies.

\subsection{Limitations of Existing Countermeasures}

Traditional approaches to combating online fraud have emphasized technological detection systems, demographic profiling of victims, legal prosecution of perpetrators, and awareness campaigns based on generic warnings \cite{anderson2008, norris2019, sarkaretal2025}. While these strategies remain essential, they offer limited support for early-stage intervention, particularly in scenarios where victims are still evaluating the legitimacy of an opportunity and have not yet incurred financial loss. Empirical evidence suggests that many individuals continue engaging with scams even after recognizing warning signs, raising a fundamental question: why do informed and educated users persist in risky interactions?

Scholarly research on job fraud remains comparatively limited, with greater attention devoted to investment fraud \cite{deevy2012}, romance scams \cite{whitty2012}, advance-fee fraud \cite{herley2012}, and online auction scams \cite{anderson2008}. Existing job fraud studies have largely adopted descriptive approaches focused on scam typologies and victim demographics, or technical perspectives emphasizing detection algorithms \cite{aleroud2017}. Far less attention has been given to systematically modeling the behavioral decision processes that drive engagement escalation and payment behavior in job scams.

\subsection{Behavioral Signals and Decision-Making}

Behavioral economics offers well-established theoretical frameworks for understanding how individuals make decisions under uncertainty, risk, and emotional pressure. Prior work demonstrates that human decision-making systematically departs from rational choice assumptions, relying instead on heuristics and exhibiting predictable cognitive biases \cite{kahneman1979, thaler1980, ariely2008}. In the context of job scams, such mechanisms can override prior knowledge or awareness, enabling fraudsters to sustain engagement and induce financial commitment.

In this study, we focus on three behavioral mechanisms that are frequently exploited in fraudulent job contexts: sunk-cost influence, where prior investments of time or effort increase commitment to a failing course of action \cite{arkes1985, thaler1980}; acute trigger cues, where loss-framed urgency and opportunity-loss pressure (fear of missing out) increase compliance under time pressure and perceived scarcity \cite{kahneman1979, cialdini2006}; and social proof, where perceived validation by others reduces skepticism under uncertainty \cite{cialdini2006}. Rather than treating these mechanisms as abstract psychological constructs, we operationalize them as measurable behavioral signals that can be evaluated in relation to observable outcomes.

As this work is a survey of job scam victimization, we adopt sunk-cost influence, acute trigger cues (loss-framed urgency and opportunity-loss pressure/FOMO), and social proof as organizing lenses because they represent a high-coverage, cross-study set of behavioral mechanisms that repeatedly appear in reported scam tactics and victim narratives. These mechanisms provide a coherent structure for synthesizing diverse findings: acute trigger cues capture urgency and opportunity-loss persuasion, social proof captures legitimacy cues and perceived peer validation under uncertainty, and sunk-cost influence explains persistence and escalation after initial engagement. While additional mechanisms (e.g., authority effects, scarcity framing, optimism bias) may also influence victim behavior, our goal is not to exhaustively model all cognitive processes, but to focus on a minimal, operationalizable set that consistently appears in job scam narratives and supports actionable intervention design.

Young adults constitute a particularly relevant population for examining these mechanisms, as they face documented susceptibility factors including financial pressure \cite{rani2016}, high unemployment rates \cite{ilo2024}, developmental factors affecting risk assessment \cite{steinberg2008, reyna2006}, and increasing exposure to sophisticated online employment scams \cite{ftc2024}.

\subsection{Contributions}

This paper makes the following contributions to human-centered security research on job scams:

\begin{itemize}
    \item \textbf{Operationalizing behavioral scam decision signals.}
    We translate behavioral economics mechanisms relevant to job scams-urgency/time-pressure cues, sunk-cost influence, and legitimacy cues (social proof)-into survey-measurable decision signals suitable for empirical analysis and intervention design.

    \item \textbf{Pathway-aware empirical analysis of job scam escalation.}
    We analyze how these signals relate to payment outcomes across common scam pathways and escalation stages, using exact small-sample inference and uncertainty-aware estimation under sparse data constraints.

    \item \textbf{Design implications for platform-level mitigation.}
    We synthesize the observed patterns into evidence-informed intervention concepts (e.g., urgency-triggered friction, sunk-cost exit prompts, targeted verification transparency, and low-cost reporting) to disrupt escalation at high-risk decision points.
\end{itemize}

The rest of the paper is organized as follows. Section II presents the methodology, including problem formulation, survey design and data collection, data preparation, and missingness handling. Section III describes the characteristics of the sample and reports the empirical results. Section IV discusses the findings, including the dominance of acute trigger cues, the limited effects of social proof, patterns of underreporting and disclosure avoidance, missingness as a behavioral signal, and study limitations. Finally, Section V concludes the paper and outlines directions for future work.

\section{Methodology}

\subsection{Problem Formulation}

We formulate job scam victimization as a behavioral signal modeling problem.
Given a set of behavioral, demographic, and contextual signals derived from self-reported scam encounters, our goal is to evaluate how specific behavioral signals relate to engagement escalation and payment behavior under realistic data scarcity constraints.
Rather than optimizing predictive accuracy, we focus on assessing the directionality, consistency, and robustness of behavioral effects across complementary analytical methods.

\subsection{Data Collection Overview}

We collected cross-sectional data using an anonymous, conditional survey administered to a university-affiliated population and its extended network. The survey captured structured responses (Likert scales, binary indicators, and numeric ranges) along with optional free-text narratives that describe scam experiences. Conditional branching ensured that participants were exposed only to questions applicable to their reported experiences, allowing targeted measurement of scam-specific behaviors while reducing survey fatigue. Detailed survey design and rationale for the platform are provided in Appendix A.

\subsection{Behavioral Signals and Operationalization}

We focus on three behavioral mechanisms frequently exploited in job scams, operationalized as measurable behavioral signals.
Table~\ref{tab:signals} summarizes the operationalization of these signals and their associated outcomes.

\begin{table}[!htbp]
\centering
\caption{Behavioral Signals and Operationalization}
\label{tab:signals}
\begin{tabular}{p{1.2cm}p{3.4cm}p{2cm}}
\hline
\textbf{Behavioral Signal} & \textbf{Measurement} & \textbf{Outcome Variables} \\
\hline
Sunk cost influence & 5-point Likert scale measuring time/effort influence & Payment (binary), payment amount \\
Acute trigger cues & Binary indicators: fear of losing opportunity, urgency & Payment (binary) \\
Social proof influence & 5-point Likert scale capturing validation cues  & Payment (binary) \\
Financial vulnerability & Composite index (income + household pressure; 0--4) & Moderation analysis \\
\hline
\end{tabular}
\end{table}

The primary outcome variable was \textbf{payment behavior}, operationalized as a binary indicator reflecting whether a participant made any financial payment during the scam encounter.
Secondary outcomes included approximate total payment amount (binned ranges) and reporting behavior (reported vs.\ not reported).

\subsection{Scam Pathways and Analytical Subsets}

Participants categorized their most recent suspicious encounter into one of three categories: \textbf{payment-based scams}, \textbf{task-based scams}, or \textbf{other/unsure}.
Payment-based and task-based scams were analyzed separately, where appropriate, to account for differences in scam structure and behavioral activation. 

Because the survey used conditional branching (e.g., scam-type and outcome-dependent question paths), different analyses required different respondent denominators. We therefore report results across nested analytical subsets defined by the survey logic: (i) the full sample of valid responses, (ii) an encounter subset restricted to participants reporting a suspicious job offer, and (iii) outcome-specific subsets for participants who made a payment and/or reported the incident. These subsets are hierarchical in the sense that each is a restriction of the previous one. Table~\ref{tab:subsets} summarizes the subsets used across evaluations.

\begin{table}[!htbp]
\centering
\caption{Analytical Subsets Used Across Evaluations}
\label{tab:subsets}
\begin{tabular}{p{2cm}p{5cm}}
\hline
\textbf{Subset} & \textbf{Description} \\
\hline
Full sample & All valid survey responses \\
Encounter subset & Participants reporting suspicious job offers \\
Payment subset & Participants who made financial payments \\
Reporting subset & Participants who reported fraud to any authority \\
\hline
\end{tabular}
\end{table}

\subsection{Data Preparation and Missingness Handling}

Data were analyzed using complete-case analysis for each specific test to avoid bias introduced by imputation in small samples. The survey employed conditional routing, which intentionally introduced structural missingness by design; such missingness was excluded from analysis. Accordingly, all missingness analyses explicitly distinguish between structural missingness arising from survey branching and item non-response, which reflects participant disengagement or sensitivity.

Beyond exclusion for estimation, missing data were examined as a potential behavioral signal. In dedicated analyses, we investigated selective non-disclosure of financial losses and emotional responses, as well as question-position effects indicative of survey fatigue and bounded rationality. We observed increasing non-response in later survey sections and systematic patterns of avoidance for sensitive items. These patterns are reported to contextualize data reliability and to assess whether non-response reflects behavioral processes rather than random noise.

\subsection{Analysis strategy}
Because the encounter-level sample is modest and several cue variables exhibit sparse counts and complete separation, our primary analysis emphasizes small-sample–appropriate exact inference and uncertainty-aware estimation. We report contingency tables and use Fisher’s exact test where cell counts are low, and we interpret effect sizes directionally under separation. To characterize stability, we complement these results with Bayesian bootstrap resampling. Additional analyses, including penalized logistic regression, propensity-score matching, random-forest feature ranking, and emotion classification, are treated as exploratory robustness checks and are not used as the basis for primary claims.

\subsection{Ethical Considerations}

The study protocol was reviewed in accordance with institutional requirements for human-subject research. Participation was voluntary and anonymous, and informed consent was obtained before data collection. No personally identifiable information was collected or stored.

\section{Results}
We first report findings from our primary exact-analysis pipeline and then summarize exploratory robustness checks to assess whether qualitative trends persist under alternative modeling assumptions.

\subsection{Sample Overview and Scam Exposure}

A total of 96 survey responses were collected through university channels. We included responses that reached the consent stage and completed the core sections required for analysis, and excluded incomplete entries that did not provide consent or lacked sufficient information to support pathway assignment and outcome coding. After this filtering, 91 participants (94.8\%) provided informed consent and completed the survey, forming the analysis sample. Among consenting respondents, 37 participants (40.7\%) reported encountering at least one suspicious job or internship offer within the past 12 months. Of these, 16 participants (43.2\%) reported making at least one payment associated with the opportunity. Table IV summarizes the sample characteristics and exposure distribution. Table~\ref{tab:sample_summary} summarizes the key sample characteristics used in subsequent analyses.

We intentionally recruited from a broader university population rather than only confirmed victims, because the study aims to model early-stage vulnerability and escalation signals, which require both exposed non-payers and payers for comparison, and because victim-only recruitment can introduce strong selection bias and limit generalizability. Despite moderate to high self-reported awareness and risk perception, nearly half of the participants encountered suspicious offers and progressed to financial loss, while formal reporting remained rare.

\begin{table}[ht]
\centering
\caption{Summary of sample and scam encounter characteristics}
\label{tab:sample_summary}
\small
\begin{tabular}{lcc}
\hline
\textbf{Characteristic} & \textbf{Count} & \textbf{Percentage} \\
\hline
Consenting respondents & 91 & 100.0\% \\
Encountered suspicious job offer & 37 & 40.7\% \\
Made payment (among encounters) & 16 & 43.2\% \\
Payment-based scams & 15 & 40.5\% \\
Task-based scams & 12 & 32.4\% \\
Reported experience & 3 & 8.1\% \\
\hline
\end{tabular}
\end{table}

\subsection{Acute Trigger Cues: Urgency/Time Pressure as a Proximal Compliance Trigger}

Acute trigger cues showed the strongest association with payment behavior in our sample, primarily through urgency/time pressure. We treat these cues as proximal compliance triggers: short-horizon persuasion signals that compress deliberation time and shift decisions toward immediate action, rather than stable traits or long-term risk preferences. In our operationalization, urgency/time pressure exhibited clear concentration among paying encounters, whereas explicit opportunity-loss (FOMO) cues were not observed in the analyzed subset. As a result, the absence of FOMO-positive cases should be interpreted as an operationalization/measurement limitation rather than evidence of no effect, and we report effect sizes directionally where separation occurs under small sample sizes.

able~\ref{tab:acute_trigger_payment} reports payment outcomes stratified by urgency/time pressure. Payment behavior was highly concentrated among encounters exhibiting urgency cues, producing complete separation in the contingency table. We therefore rely on Fisher’s exact test for inference and interpret effect sizes directionally rather than as stable population estimates. The association between urgency cues and payment is statistically detectable in this sample (Fisher’s exact test, $p=0.005$), but the resulting non-finite odds ratio reflects small-sample separation rather than an infinitely large real-world effect.

In contrast, FOMO/opportunity-loss pressure could not be reliably evaluated under the current extraction procedure. Specifically, no encounters were classified as FOMO-positive(\texttt{fomo\_used}=1), and all observed payments occurred in the FOMO-negative group (4/37; 10.81\%). Because the contingency table collapses to a single observed row, a standard 2\(\times\)2 Fisher’s test cannot be computed, and the dataset does not permit inferential claims about FOMO’s relationship with payment escalation in this subsample. This absence of detected FOMO-positive cases may reflect either (i) limited prevalence of explicit FOMO tactics in the encounter narratives captured here, or (ii) under-detection due to conservative cue-identification rules. 

Accordingly, while urgency emerges as an empirically supported acute trigger for escalation, FOMO effects remain indeterminate and require improved operationalization (e.g., expanded lexicons or manual annotation) before drawing conclusions. From an intervention perspective, these results suggest urgency cues may be a high-yield target for early warnings and friction mechanisms designed to slow down decision-making during scam interactions. One explanation is that opportunity-loss framing may be expressed implicitly (e.g., “slots filling fast”) or through culturally localized phrasing not captured by our survey operationalization, warranting refined measurement and annotation in future work. 

\begin{table}[t]
\centering
\caption{Payment behavior stratified by acute trigger cues (\(n=37\))}
\label{tab:acute_trigger_payment}
\small
\begin{tabular}{lccc}
\hline
\multicolumn{4}{l}{\textbf{Urgency / Time Pressure}} \\
\hline
\textbf{Cue Present} & \textbf{Not Paid (0)} & \textbf{Paid (1)} & \textbf{Payment Rate} \\
\hline
No (0)  & 26 & 0 & 0.00\% \\
Yes (1) & 7  & 4 & 36.36\% \\
\hline
\multicolumn{4}{l}{} \\
\hline
\multicolumn{4}{l}{\textbf{FOMO / Opportunity-Loss Pressure}} \\
\hline
\textbf{Cue Present} & \textbf{Not Paid (0)} & \textbf{Paid (1)} & \textbf{Payment Rate} \\
\hline
No (0)  & 33 & 4 & 10.81\% \\
Yes (1) & 0  & 0 & -- \\
\hline
\end{tabular}
\end{table}

\subsection{Sunk Cost Influence and Escalation of Commitment}

Sunk cost influence exhibited a consistent but weaker relationship with payment behavior. Participants reporting a stronger influence of previously invested time or effort were more likely to make payments. Table~\ref{tab:sunk_cost_payment} summarizes payment rates by sunk cost influence level.

\begin{table}[ht]
\centering
\caption{Payment rates by sunk cost influence level}
\label{tab:sunk_cost_payment}
\small
\begin{tabular}{lcc}
\hline
\textbf{Sunk Cost Influence} & \textbf{Payment Rate} & \textbf{N} \\
\hline
Low (1--2) & 31.8\% & 22 \\
High (3--5) & 60.0\% & 15 \\
\hline
\end{tabular}
\end{table}

Participants with high sunk cost influence were 1.9$\times$ more likely to make payments than those with low influence. Although Fisher’s Exact Test did not reach conventional significance ($p = 0.107$), Bayesian estimation yielded a median odds ratio of 1.60, with 92.3\% posterior probability mass above OR $= 1$, indicating consistent evidence for a positive effect under sampling uncertainty.

\subsection{Financial Vulnerability and Conditional Risk}

Payment behavior was concentrated within a small subgroup characterized by both elevated financial vulnerability and acute trigger cues. All observed payments occurred among financially vulnerable participants who endorsed at least one acute trigger cues. No payments were observed outside this intersection.

Due to complete or near-complete separation, interaction terms could not be reliably estimated using logistic regression. The descriptive concentration pattern itself, however, highlights a high-risk subgroup with clear implications for targeted intervention.

\subsection{Social Proof Effects}

Self-reported social proof influence did not independently predict payment behavior. Median social proof influence scores were low across both payers and non-payers, and non-parametric comparisons revealed no statistically significant differences (p = 0.431). Because social proof influence is measured via self-report and is socially evaluative, responses may be affected by social desirability and self-presentation biases, potentially attenuating observed associations. Exploratory cross-tabulations suggested that combined legitimacy cues (professional appearance together with social proof elements) may increase payment likelihood. However, extreme cell sparsity limits inference, and these results are treated as hypothesis-generating and deferred to the appendix.
\subsubsection{Social Proof and Legitimacy Cues}
In our sample, social proof cues did not show a statistically detectable standalone association with payment behavior under exact testing. This null result should not be interpreted as evidence that social proof is irrelevant in job scams; rather, it suggests that the influence of social validation cues may be context-dependent and may operate through co-occurrence with other legitimacy signals (e.g., professional appearance, branding, or recruiter credibility). Accordingly, we treat social proof primarily as a design-relevant risk context that may amplify persuasion when bundled with payment demands, rather than as a consistently separable predictor in this dataset.

\subsection{Reporting Behavior and Disclosure Avoidance}

Only 3 out of 37 scam encounter respondents (8.1\%) reported their experience to any authority or platform. Reporting rates did not vary significantly by loss magnitude, indicating that perceived procedural and psychological costs dominate reporting decisions. At the same time, disclosure avoidance may still scale with loss magnitude in terms of the actual amount paid, as larger losses may increase shame, self-blame, or perceived reputational cost even when they do not translate into higher formal reporting rates \cite{Breen2022, Correia2022, Sarkar2024}. Among payment-based scam victims, 73.3\% did not disclose exact financial loss amounts. This asymmetry between emotional disclosure and financial non-disclosure is consistent with avoidance of loss salience and self-concept protection.

\subsection{Emotional Signals in Victim Narratives (NLP Analysis)}

To complement the structured survey responses, we examined the affective signals present in respondents' free-text scam narratives. Of the 96 respondents, 35 (36.5\%) provided narrative text of sufficient length for qualitative or NLP-assisted analysis. Table~\ref{tab:emotion_distribution} summarizes the dominant emotion categories observed in these narratives.

\begin{table}[ht]
\centering
\caption{Emotion distribution in scam-related narratives (n=35)}
\label{tab:emotion_distribution}
\small
\begin{tabular}{lcc}
\hline
\textbf{Emotion} & \textbf{Count} & \textbf{Percentage} \\
\hline
Fear & 15 & 42.86\% \\
Anger & 3 & 8.57\% \\
Joy & 1 & 2.86\% \\
Other/Mixed & 10 & 28.57\% \\
No Emotion & 6 & 17.14\% \\
\hline
\end{tabular}
\end{table}

Fear emerged as the most frequently expressed dominant emotion (42.86\%), indicating that scam encounters are commonly framed in terms of uncertainty, perceived threat, and anxiety about consequences. Anger was comparatively less frequent (8.57\%), which may reflect that the immediate reaction in many cases is dominated by fear, shock, or self-doubt rather than moral outrage. Positive affect was rare (joy: 2.86\%), consistent with the inherently adverse nature of scam experiences. A substantial share of narratives exhibited either mixed affect (Other/Mixed: 28.57\%) or relatively affect-neutral descriptions (No Emotion: 17.14\%), suggesting heterogeneity in how victims narrate harm, ranging from emotionally charged accounts to detached or factual recounting.

We interpret these labels as situational signals expressed in victim narratives rather than as indicators of stable personality traits. In addition, due to computational and environmental constraints, we treat the emotion distribution as a descriptive summary rather than a full transformer-based modeling result, and avoid inferring fine-grained emotional dynamics beyond the observed distribution.

\subsection{Summary of Findings}
Overall, the results indicate that job scam payments can be interpreted using a
two-stage persuasion structure: (i) escalation mechanisms that increase continued
engagement over time, and (ii) loss-framed triggers that more directly precipitate
payment decisions. Specifically:

\begin{itemize}
    \item Payment decisions are most strongly associated with urgency/time-pressure cues. In contrast, fear of missing out (FOMO) could not be evaluated in this sample because no encounters were classified as FOMO-positive.
    \item Sunk-cost dynamics contribute to escalation by increasing continued engagement after prior investments of time and effort, but appear secondary to acute loss framing in explaining payment behavior.
    \item Financial vulnerability amplifies susceptibility to these triggers, indicating that psychological pressure is more consequential under constrained economic conditions.
    \item Social proof alone shows limited independent explanatory power in this sample, although it may operate in combination with other credibility cues.
    \item Formal reporting remains extremely rare, and observed reporting patterns appear largely independent of loss magnitude, highlighting persistent barriers and stigma in help-seeking.
\end{itemize}

Importantly, the ``acute trigger cues'' constructs examined in this study refer to
loss-framed persuasion mechanisms and emotional responses to potential losses
(e.g., urgency and fear of missing an opportunity), rather than a stable
risk-aversion trait. In this framing, the decision pressure arises from the perception
that \emph{not acting now may produce a loss}, even when the objective stakes may be small.

\section{Discussion}

This study examined the behavioral economic mechanisms underlying payment decisions in online job scams among Indian university students. We tested whether acute trigger cues, sunk cost influence, and social proof mechanisms are associated with scam escalation and payment behavior, and evaluated how financial vulnerability and reporting frictions shape victim outcomes.

\subsection{Dominance of Acute Trigger Cues in Payment Decisions}

Our findings indicate that short-horizon trigger cues are among the most salient correlates of payment decisions in job scams. In particular, urgency/time pressure was strongly associated with financial commitment in the encounter subset (Table~V). Participants exposed to urgency/time pressure cues exhibited a 36.4\% payment rate (4/11), compared to 0\% (0/26) among those without such cues (Fisher’s exact test, $p = 0.005$). In contrast, fear of losing an opportunity (opportunity-loss pressure or FOMO) could not be evaluated in this sample because no encounters were classified as FOMO-positive, preventing effect estimation for that cue (Table~V). Due to complete separation in the urgency contingency table, effect-size estimates (e.g., odds ratios) should be interpreted directionally rather than as stable population parameters.

Importantly, these acute trigger cues may also interact with financial vulnerability. In this sample, payments were observed only among financially vulnerable participants who also exhibited urgency/time-pressure cues, whereas no payments were observed in the other observed combinations. While this pattern is suggestive, the small sample size and separation preclude stable interaction-effect estimation. Nonetheless, the concentration of payments in the vulnerable + urgency subgroup highlights a practically meaningful risk cluster that may be especially intervention-relevant.

\subsection{Sunk Cost as a Secondary Mechanism}

Evidence for sunk cost effects was suggestive but weaker. Participants reporting higher sunk cost influence exhibited higher payment rates (60.0\% vs.\ 31.8\%), corresponding to a risk ratio of 1.89, though this association did not reach conventional significance thresholds ($p = 0.107$). However, multiple robustness checks converged toward a positive effect: Bayesian estimation indicated a 92.3\% posterior probability that sunk cost influence increased payment odds, and propensity score matching produced a +33.3 percentage point average treatment effect.

Taken together, these findings suggest that sunk cost operates as a background escalation mechanism rather than a proximal trigger. Prior investments of time and effort appear to increase susceptibility \cite{Feldman2018, Brockner1992}, but payment decisions are more directly activated by immediate trigger cues such as urgency/time pressure \cite{Naidoo2025, Carter2023, Pratama2024}, and potentially opportunity-loss (FOMO) framing, although the latter could not be evaluated reliably in our encounter subset under the current operationalization.

\subsection{Null and Exploratory Findings for Social Proof}
Contrary to theoretical expectations, self-reported social proof influence did not significantly predict payment behavior (p = 0.431), nor was the effect moderated by age. This pattern may reflect characteristics of the study population, such as higher digital literacy and familiarity with manufactured engagement cues among university students, which could attenuate reliance on superficial validation signals. An additional explanation is measurement attenuation: self-reported susceptibility to social proof may be underreported due to social desirability and self-presentation concerns, particularly in scam victimization contexts. Self-reported endorsement of social influence may be conservative due to social desirability and self-attribution biases, which can attenuate observed effects.

 Exploratory descriptive patterns suggest that the conjunction of professional appearance and social validation artifacts may coincide with higher payment propensity, motivating further study with larger samples. Due to extreme cell sparsity, this pattern must be interpreted cautiously and treated as hypothesis-generating rather than confirmatory. Overall, social proof appears to play a weaker and more context-dependent role than acute trigger cues (urgency and opportunity-loss pressure) and escalation mechanisms (sunk cost) in this setting.

\paragraph{Implications for legitimacy-oriented protections.}
Given the weak standalone association for social proof cues, broad ``anti-testimonial'' warnings may be less effective than context-aware verification affordances. Platforms can instead prioritize targeted friction and transparency mechanisms when professional appearance cues and social validation artifacts (e.g., testimonials, earnings screenshots) co-occur with payment requests, as these bundles may elevate perceived legitimacy even when individual cues are insufficient on their own.

\subsection{Underreporting, Disclosure Avoidance, and Missingness as Signal}

The observed 91.9\% non-reporting rate underscores the magnitude of underreporting in job scam victimization. Reporting behavior was not significantly associated with loss magnitude, indicating that practical barriers such as transaction costs, time, effort, emotional burden, and low expectations of resolution may discourage formal help-seeking. In addition, underreporting can reflect social censure and fear of ridicule, where victims anticipate blame, embarrassment, or judgment when disclosing fraud experiences. This aligns with prior research documenting systemic barriers and stigma in fraud reporting.

Financial loss disclosure exhibited substantial non-response: 73.3\% of payment-based victims declined to report amounts paid. This pattern is consistent with selective non-disclosure for financially sensitive details, and it may be amplified when the actual amount paid increases the perceived psychological and reputational cost of disclosure, even if it does not translate into higher formal reporting rates. In contrast, emotional disclosure rates were higher, suggesting that financial losses are uniquely sensitive and may be harder to disclose.

Missingness increased across later survey questions ($\rho = 0.346$), which is consistent with survey fatigue and time loss as primary explanations for incomplete responses. Given this, missing data should be treated conservatively as a data-quality limitation rather than over-interpreted as evidence of bounded rationality or cognitive depletion. At the same time, the concentration of non-response within financially sensitive items suggests that missingness may reflect a combination of fatigue and sensitivity-related reluctance, motivating future work using shorter instruments or alternative measurement strategies to reduce respondent burden while improving disclosure reliability.

\subsection{Implications for System Design}

Our findings suggest that effective job-scam mitigation should target loss-framed decision points in scam progression, rather than relying solely on broad awareness or education. In particular, urgency/time-pressure cues appear to function as proximal compliance triggers that compress deliberation and accelerate payment decisions. Accordingly, just-in-time interventions that introduce lightweight friction at payment moments and reframe urgency-driven compliance in terms of avoidable financial and security loss may be especially high-yield. In addition, escalation patterns consistent with sunk-cost influence motivate exit prompts that make prior investment (time, steps, interaction depth) explicit and decouple it from future payment decisions. By contrast, social-proof mitigation likely requires more context-sensitive deployment, as credibility cues may be most harmful when bundled with professional appearance signals and payment demands. Finally, the low reporting rate observed in our sample motivates interventions that reduce the transaction costs of reporting and provide feedback that reinforces prosocial impact.

We translate these observed association patterns into evidence-informed intervention concepts that map trigger signals to deployable design patterns (Table~\ref{tab:interventions}). These concepts should be interpreted as design hypotheses rather than validated deployable detection features, since our signals are measured via retrospective survey reports and evaluated under small-$n$ constraints.

\begin{table*}[t]
\centering
\caption{Evidence-informed intervention concepts mapped to observed trigger signals, behavioral mechanisms, and deployable design patterns.}
\label{tab:interventions}
\small
\setlength{\tabcolsep}{6pt}
\renewcommand{\arraystretch}{1.25}
\begin{tabular}{p{2.7cm} p{3.2cm} p{4.2cm} p{5.6cm}}
\hline
\textbf{Intervention} & \textbf{Observed Trigger Signal} & \textbf{Target Mechanism} & \textbf{Design Pattern / Action} \\
\hline

\textbf{I1. Urgency warning} &
Urgency/time-pressure cues near payment request &
Proximal compliance under time compression &
Just-in-time friction before payment actions; loss-reframing prompt emphasizing financial/security loss; quick verification checklist (employer identity, official domain, no-fee rule). \\

\textbf{I2. Sunk-cost exit prompt} &
High interaction depth / repeated steps / prolonged thread &
Escalation of commitment from prior effort (sunk-cost influence) &
Progress transparency (time/steps invested); explicit decoupling (``past effort does not justify paying''); one-tap safe exit (``Stop and verify''). \\

\textbf{I3. Social proof verification friction} &
Professional appearance cues and/or social validation artifacts (e.g., testimonials, screenshots) &
Trust inflation via credibility bundling (social proof $\times$ professional appearance) &
Verification transparency (what badges certify); counter-social-proof nudge (``screenshots are easy to fake''); stronger friction when credibility cues co-occur with payment requests. \\

\textbf{I4. Risk-tiered warnings} &
Vulnerability indicators co-occurring with urgency cues &
Heightened compliance susceptibility under vulnerability $\times$ pressure &
Tiered warning intensity for urgency + payment contexts; pair warnings with vetted alternative job resources; avoid invasive profiling by using conservative triggers. \\

\textbf{I5. One-click reporting} &
Low reporting uptake post-incident &
High transaction costs + psychosocial barriers to reporting &
One-click reporting with pre-filled context (account/link/payment handle); optional anonymity; impact feedback (``your report helps protect others''). \\

\hline
\end{tabular}
\end{table*}

\subsection{Limitations}

This study has several limitations that constrain inference, generalizability, and effect size interpretation. These limitations are important for contextualizing the findings and guiding future research.

\subsubsection{Sampling and Generalizability Constraints}

\textbf{Small sample size.}  
The total sample comprised 96 respondents, of whom only 37 (38.5\%) reported scam encounters and 16 (16.7\%) made payments. This severely limits statistical power, particularly for stratified and interaction analyses, where cell counts dropped to as low as $n=1$--$4$. As a result, effect estimates are unstable and confidence intervals are wide (e.g., sunk cost OR 95\% CI [0.82, 3.80]). Bayesian analyses allowed probabilistic interpretation despite non-significance, but posterior distributions remain broad. The payer subsample is especially constrained: significance for some trigger cues is driven by perfect or near-perfect separation (zero cells), and effect estimates should therefore be interpreted cautiously and validated in larger samples.

\textbf{Convenience sampling from a single institution.}  
Participants were recruited from the IIIT Hyderabad community, yielding a relatively homogeneous population that is young (mean age 21.7 years), highly educated, and technologically proficient. This group likely possesses greater digital literacy and scam awareness than the general job-seeking population, potentially attenuating social proof effects and underestimating vulnerability among older, less educated, or rural job seekers. Cultural factors specific to India, such as stigma around victimization, family financial interdependence, and acceptance of upfront training fees, may further limit generalizability to other national contexts.

\textbf{Self-selection and non-response bias.}  
Five respondents (5.2\%) explicitly refused consent, indicating unit non-response. More critically, individuals experiencing severe financial or psychological harm may avoid participation due to embarrassment, avoidance coping, or distrust. The observed 8.1\% reporting rate suggests widespread reluctance to disclose victimization, which may extend to survey participation itself. Consequently, payment amounts, emotional impact, and escalation patterns observed here likely represent lower bounds. Conversely, participants motivated to warn others may be overrepresented, inflating awareness and recognition confidence.

\textbf{University affiliation bias.}  
Recruitment through institutional mailing lists and student groups may disproportionately attract respondents with higher institutional trust and civic engagement, further limiting representativeness.

\subsubsection{Measurement Validity and Self-Report Limitations}

\textbf{Retrospective self-report.}  
All measures rely on retrospective recall of scam encounters within a 12-month window. Recall bias, post-hoc rationalization, and emotion reconstruction may distort reported payment amounts, decision rationales, and emotional states. Social desirability pressures may suppress embarrassment reporting while inflating perceived scam recognition ability.

\textbf{Payment amount accuracy.}  
Among payment-based victims, 73.3\% declined to disclose payment amounts, producing substantial missing data bias. Among the small disclosure subset ($n=4$), amounts ranged widely (Rs.\,1{,}250 to Rs.\,75{,}000), but verification was not possible. The use of categorical ranges and midpoint imputation introduces additional measurement error, particularly for the open-ended upper category (``Rs.\,50{,}000+'').

\textbf{Construct operationalization.}  
Key constructs (urgency/time pressure, fear of losing an opportunity, sunk-cost influence, and social proof) were measured using single-item Likert scales or binary checklist indicators rather than validated multi-item instruments. While appropriate for exploratory analysis under survey length constraints, single-item measures may exhibit lower reliability and limited construct coverage.

\textbf{Emotional measurement constraints.}  
Structured emotion scales capture retrospective emotional impact rather than in-the-moment affect during scam encounters. Consequently, temporal directionality cannot be established: emotions may precede, follow, or co-evolve with behavioral outcomes. The NLP-based emotion analysis partially addresses this gap but is limited by a small narrative subsample.

\paragraph{Interpreting missingness.}
Because our study uses retrospective self-reports, some items exhibit nonresponse that may reflect a mix of factors, including cognitive burden, uncertainty, recall limitations, and sensitivity around self-blame or embarrassment. We therefore avoid treating missingness as direct evidence of a specific psychological mechanism. Instead, we treat nonresponse as a potentially informative measurement artifact that can bias estimates if ignored and may itself be relevant for intervention design (e.g., reducing reporting and disclosure friction). Accordingly, we distinguish structural missingness induced by scam pathway branching from item non response, and we perform sensitivity analyses to assess whether key qualitative patterns persist under alternative assumptions.

\subsubsection{Statistical and Methodological Constraints}

\textbf{Perfect and quasi-complete separation.}  
Several logistic regression models, particularly for acute trigger cues such as urgency/time pressure and fear of losing an opportunity, exhibited perfect or near-perfect prediction due to zero cells in the contingency tables. This produced extremely large or undefined odds ratios (e.g., OR = 273.0 for urgency pressure) and numerically unstable coefficients. While these patterns indicate strong directional associations within this sample, the magnitudes should not be interpreted as stable population parameters.

\textbf{Multiple comparisons.}  
Numerous statistical tests were conducted across multiple hypotheses without global family-wise error correction. Although false discovery rate correction was applied in selected post-hoc analyses, uncorrected testing increases Type I error risk. Some nominally significant results may therefore represent false positives.

\textbf{Assumption violations.}  
Severe non-normality, skewed payment distributions, and sparse contingency tables necessitated non-parametric tests (Fisher's Exact, Mann--Whitney U, Kruskal--Wallis). While robust, these tests have lower power and test rank-based rather than mean-based hypotheses, potentially obscuring subtle effects.

\textbf{Propensity score matching limitations.}  
Propensity matching for sunk cost analyses relied on a small number of matched pairs and coarse dichotomization, limiting balance quality. Unobserved confounders, such as personality traits or prior scam exposure, remain uncontrolled.

\subsubsection{Temporal and Causal Inference Limitations}

\textbf{Cross-sectional design.}  
All data were collected at a single time point, precluding causal inference. Although question wording implies temporal ordering (e.g., prior investment influencing later payment), reverse causation and third-variable confounding cannot be ruled out.

\textbf{Survivor bias.}  
The study captures only individuals who recognized suspicious encounters, remained able and willing to participate, and survived potential losses. Victims who paid substantial amounts without realizing fraud or experienced severe downstream consequences are structurally excluded.

\textbf{Temporal ambiguity in narratives.}  
Emotions extracted from free-text narratives may reflect in-the-moment reactions, retrospective appraisal, or current anxiety about future scams. Mediation analyses assume temporal ordering that cannot be verified with cross-sectional text data.

\subsubsection{Platform and Context-Specific Constraints}

\textbf{Platform distribution bias.}  
The dominance of WhatsApp/Telegram encounters (51.4\%) may reflect communication norms among Indian students rather than universal scam distributions. Other populations may experience higher exposure via email, phone calls, or job portals, which were underrepresented here.

\textbf{Scam type heterogeneity.}  
Payment-based, task-based, and ambiguous scam types were analyzed jointly in most tests due to sample size constraints. These scam types likely operate through distinct psychological mechanisms, which could not be reliably disentangled.

\textbf{Missing data mechanisms.}  
Although missingness was analyzed as a behavioral signal, structural missingness introduced by survey branching complicates interpretation and limits separation of avoidance-driven non-response from design-induced absence.

\textbf{NLP model limitations.}  
Emotion classification relied on pre-trained models without domain-specific validation. The small narrative subsample and potential expressive bias (emotionally affected respondents writing longer narratives) further limit inference. Consequently, NLP findings should be interpreted as exploratory.

\section{Conclusion and Future Work}

This study examined behavioral mechanisms associated with job scam victimization, with a focus on how escalation dynamics and short-horizon trigger cues shape payment decisions. Using a mixed-methods approach combining statistical analysis, interpretable feature-importance analyses, Bayesian estimation, and computational emotion analysis, we show that job scams can leverage predictable patterns in attention, emotion, and decision pressure rather than relying solely on technical ignorance or lack of awareness.

Across analyses, payment behavior was most strongly associated with acute trigger cues, with urgency/time pressure emerging as the clearest proximal compliance signal in scam escalation. While opportunity-loss (FOMO) framing is theoretically central to loss-based persuasion, it was not observable under our current cue-identification procedure in the encounter subset, motivating refined measurement in future work. These trigger cues were especially consequential under financial vulnerability, revealing an empirically distinct high-risk subgroup that is both intervention-relevant and policy-relevant. In contrast, sunk-cost dynamics appear to function primarily as a background escalation mechanism: prior investments of time and effort may increase continued engagement, but payment decisions are most directly precipitated by immediate trigger cues. Social proof exhibited limited independent explanatory power in this sample, suggesting that its influence may be context-dependent and most effective when bundled with additional credibility signals.

Beyond payment outcomes, formal reporting was extremely rare. Non-response increased in later survey sections, consistent with survey fatigue and time loss as primary contributors to missing data, while non-disclosure was especially concentrated in financially sensitive questions. Taken together, these patterns suggest that underreporting and incomplete disclosure reflect both practical barriers and social factors such as stigma, social censure, and fear of ridicule, which can suppress help-seeking even when victimization occurs.

By grounding job scam victimization in operationalizable behavioral cues, this work contributes a principled human-centered framework for identifying where in the scam trajectory interventions may be most effective. Rather than treating victims as irrational or careless, our results highlight systematic pressures that can be anticipated and mitigated through platform design choices and targeted warning mechanisms.

Future work should prioritize longitudinal and experimental designs that capture scam encounters in real time and support causal inference. Ecological momentary assessment and randomized experiments testing behaviorally informed interventions, such as cooling-off periods, friction-based prompts, and context-sensitive warnings, would directly evaluate whether reducing urgency and opportunity-loss pressure lowers payment rates.

Replication with larger and more demographically diverse samples is necessary to assess generalizability beyond a university population and to obtain stable estimates for sparse interaction patterns and loss magnitude effects. Cross-cultural studies could further clarify how stigma, financial norms, and platform usage moderate vulnerability.

Finally, partnerships with job platforms and payment providers could enable analysis of objective behavioral traces, such as interaction timing, message cadence, and escalation patterns, to complement self-report measures. Evaluating the feasibility and cost-effectiveness of adaptive, machine learning, assisted interventions and coordinating response mechanisms across platforms, financial institutions, and regulators remains an important direction for scalable deployment.

\FloatBarrier

\appendices
\section{Survey Design and Instrumentation}

This appendix documents the survey design, instrumentation, routing logic, and ethical safeguards to support transparency and replicability. The main paper focuses on behavioral signal modeling and empirical findings.

\subsection{Survey Platform Selection}

Google Forms was selected based on accessibility, anonymity support, and conditional routing capabilities. Table~\ref{tab:platform_rationale} summarizes the selection rationale.

\begin{table}[!t]
\centering
\caption{Survey Platform Selection Rationale}
\label{tab:platform_rationale}
\small
\begin{tabular}{p{2.2cm}p{3.8cm}}
\hline
\textbf{Criterion} & \textbf{Justification} \\
\hline
Accessibility & No installation required \\
Familiarity & Widely used by students \\
Cost & Free, unlimited responses \\
Privacy & No IP/email collection \\
Conditional logic & Supports branching flows \\
Validation & Built-in numeric checks \\
\hline
\end{tabular}
\end{table}

\subsection{Anonymous Survey Design}

The survey was fully anonymous to reduce stigma and encourage disclosure of sensitive experiences. Table~\ref{tab:anonymity_rationale} summarizes the design rationale.

\begin{table}[!t]
\centering
\caption{Rationale for Anonymous Survey Design}
\label{tab:anonymity_rationale}
\small
\begin{tabular}{p{2.8cm}p{3.4cm}}
\hline
\textbf{Factor} & \textbf{Benefit} \\
\hline
Social stigma & Reduces fear of judgment \\
Response honesty & Limits desirability bias \\
Financial disclosure & Enables loss reporting \\
Psychological safety & Supports bias admission \\
Participation & Improves response rates \\
\hline
\end{tabular}
\end{table}

\subsection{Target Population}

The target population consisted of members of a university community and extended networks, including students, staff, faculty, alumni, and their contacts. This population was selected due to high exposure to online job opportunities and early-career employment risks.

\subsection{Survey Structure and Routing}

The survey employed conditional branching based on consent, encounter experience, and scam type. Table~\ref{tab:survey_sections} summarizes the structure.

\begin{table}[!t]
\centering
\caption{Survey Sections and Routing Overview}
\label{tab:survey_sections}
\small
\begin{tabular}{p{2.8cm}p{3.6cm}}
\hline
\textbf{Section} & \textbf{Purpose} \\
\hline
Consent & Informed participation \\
Demographics & Background variables \\
Awareness & Scam familiarity \\
Encounter details & Suspicious offer data \\
Payment-based & Fee-related scams \\
Task-based & Gamified scams \\
Aftermath & Emotions and reporting \\
Safety & Risk perception \\
\hline
\end{tabular}
\end{table}

\subsection{Demographic and Socioeconomic Variables}

Demographic and socioeconomic variables were used as controls and moderators. Table~\ref{tab:demographic_vars} summarizes these measures.

\begin{table}[!t]
\centering
\caption{Demographic Variables and Analytical Role}
\label{tab:demographic_vars}
\small
\begin{tabular}{p{2.2cm}p{3.2cm}}
\hline
\textbf{Variable} & \textbf{Purpose} \\
\hline
Age group & Developmental effects \\
Education level & Skill protection \\
Income range & Financial vulnerability \\
Household status & Economic pressure \\
Job status & Scam exposure risk \\
Geography & Regional control \\
\hline
\end{tabular}
\end{table}

\subsection{Encounter-Level Variables}

Table~\ref{tab:encounter_vars} lists variables capturing behavioral signals during scam encounters.

\begin{table}[!t]
\centering
\caption{Encounter-Level Variables}
\label{tab:encounter_vars}
\small
\begin{tabular}{p{2.8cm}p{3.2cm}}
\hline
\textbf{Variable} & \textbf{Purpose} \\
\hline
Contact platform & Exposure channel \\
Offer type & Scam typology \\
Warning signs & Awareness–action gap \\
Sunk cost influence & Escalation signal \\
Social proof influence & Validation susceptibility \\
Engagement level & Outcome severity \\
\hline
\end{tabular}
\end{table}

\subsection{Scam-Type-Specific Variables}

The variables used to model payment-based and task-based job scam pathways are summarized in Table~\ref{tab:payment_vars} and Table~\ref{tab:task_vars}, respectively.

\begin{table}[!t]
\centering
\caption{Payment-Based Scam Variables}
\label{tab:payment_vars}
\small
\begin{tabular}{p{2.8cm}p{3.2cm}}
\hline
\textbf{Variable} & \textbf{Purpose} \\
\hline
Initial payment & Commitment anchor \\
Total payments & Escalation depth \\
Total amount paid & Financial harm \\
Urgency / fear cues & Loss aversion \\
Payment method & Traceability \\
\hline
\end{tabular}
\end{table}

\begin{table}[!t]
\centering
\caption{Task-Based Scam Variables}
\label{tab:task_vars}
\small
\begin{tabular}{p{2.8cm}p{3.cm}}
\hline
\textbf{Variable} & \textbf{Purpose} \\
\hline
Initial reward & Reciprocity cue \\
Tasks completed & Engagement depth \\
Gamification cues & Behavioral activation \\
Investment request & Bait-switch signal \\
Total amount paid & Financial harm \\
\hline
\end{tabular}
\end{table}






\section{Supplementary Results and Robustness Analyses}

This appendix reports additional statistical results, robustness checks, and diagnostic analyses that support the main findings but were omitted from the Results section to preserve narrative clarity and space constraints in the two-column IEEE format.


\subsection{Behavioral Signal Evaluation}

Associations between behavioral signals and payment behavior were evaluated using complementary analytical methods.
Table~\ref{tab:methods} maps research objectives to the analytical techniques employed.

\begin{table}[!htbp]
\centering
\caption{Analytical Methods by Research Objective}
\label{tab:methods}
\begin{tabular}{p{3.4cm}p{4cm}}
\hline
\textbf{Objective} & \textbf{Methods Used} \\
\hline
Behavioral signal association & Fisher’s Exact Test, odds ratios, risk ratios \\
Monotonic relationships & Mann--Whitney U test, Spearman correlation \\
Moderation by vulnerability & Penalized logistic regression with interaction terms \\
Robustness under data scarcity & Bayesian bootstrap (5,000 iterations), threshold sensitivity analysis \\
Relative signal importance & Random Forest feature importance (depth-limited) \\
\hline
\end{tabular}
\end{table}

Bayesian bootstrap estimation generated posterior distributions of odds ratios to enable probabilistic assessment of effect direction under uncertainty.
Sensitivity analyses tested robustness to threshold choices by varying cutpoints used to dichotomize behavioral signals and re-evaluating associations across thresholds.

\subsection{Exploratory Narrative Analysis}

Optional free-text narratives describing scam experiences were analyzed using a pre-trained DistilBERT-based emotion classifier, following Ekman-style categories. This analysis was explicitly exploratory and aimed to assess whether emotional tone relates to payment amount, reporting behavior, or scam type.
All limitations related to model domain mismatch, temporal ambiguity, and reduced subsample size are acknowledged.

\subsection{Descriptive Cross-Tabulations}

\subsubsection*{Platform by Engagement Outcome}

Table~\ref{tab:platform_engagement_appendix} presents engagement rates by initial contact platform. While platform effects were not central to hypothesis testing, these patterns contextualize exposure channels. The high engagement rate on social media platforms should be interpreted cautiously due to small cell sizes.

\begin{table}[!t]
\centering
\caption{Engagement Rates by Initial Contact Platform}
\label{tab:platform_engagement_appendix}
\small
\begin{tabular}{lcc}
\hline
\textbf{Platform} & \textbf{Engaged (\%)} & \textbf{N} \\
\hline
WhatsApp / Telegram & 26.3 & 19 \\
Email & 33.3 & 6 \\
LinkedIn & 75.0 & 4 \\
Instagram / Facebook & 100.0 & 3 \\
Phone / SMS & 50.0 & 2 \\
\hline
\end{tabular}
\end{table}

\subsection{Sunk Cost Sensitivity Analysis}

To test the robustness of the sunk cost effect, multiple threshold definitions were evaluated. Results are summarized in Table~\ref{tab:sunk_cost_sensitivity_appendix}. The effect is strongest at moderate sunk cost thresholds, suggesting escalation begins early rather than only at extreme commitment levels.

\begin{table}[!t]
\centering
\caption{Sensitivity of Sunk Cost Effect to Threshold Definition}
\label{tab:sunk_cost_sensitivity_appendix}
\small
\begin{tabular}{lccc}
\hline
\textbf{Threshold} & \textbf{Odds Ratio} & \textbf{p-value} & \textbf{N (High)} \\
\hline
$\geq$2.5 & 3.21 & 0.107 & 15 \\
$\geq$3.0 & 3.21 & 0.107 & 15 \\
$\geq$3.5 & 1.36 & 1.000 & 14 \\
$\geq$4.0 & 1.36 & 1.000 & 14 \\
\hline
\end{tabular}
\end{table}

\subsection{Social Proof Robustness Checks}

\subsubsection*{Social Proof by Age Group}

Table~\ref{tab:social_proof_age_appendix} reports correlations between social proof influence and payment stratified by age. No age-based moderation of social proof susceptibility was observed.

\begin{table}[!t]
\centering
\caption{Social Proof Influence by Age Group}
\label{tab:social_proof_age_appendix}
\small
\begin{tabular}{lcc}
\hline
\textbf{Age Group} & $\boldsymbol{\rho}$ & \textbf{p-value} \\
\hline
$\leq$24 years & 0.040 & 0.826 \\
$>$24 years & -0.500 & 0.500 \\
\hline
\end{tabular}
\end{table}

\subsection{Emotion Analysis Supplement}

\subsubsection*{Emotion Distribution by Scam Type}

Table~\ref{tab:emotion_scam_appendix} reports the distribution of dominant emotions across scam categories using row percentages. Given the small cell sizes within emotion--scam combinations, we interpret these patterns descriptively rather than inferentially.

\begin{table}[!t]
\centering
\caption{Emotion Distribution by Scam Type (Row \%)}
\label{tab:emotion_scam_appendix}
\small
\begin{tabular}{lccc}
\hline
\textbf{Emotion} & \textbf{Payment} & \textbf{Task} & \textbf{Other} \\
\hline
Fear & 35.7 & 50.0 & 22.2 \\
Surprise & 21.4 & 0.0 & 33.3 \\
Sadness & 14.3 & 8.3 & 22.2 \\
Joy & 0.0 & 33.3 & 0.0 \\
Neutral & 21.4 & 8.3 & 0.0 \\
Anger & 7.1 & 0.0 & 22.2 \\
\hline
\end{tabular}
\end{table}

Several descriptive patterns emerge. Fear is the most prominent emotion overall and appears most strongly in task-based scams (50.0\%), consistent with the uncertainty and escalating engagement structure typical of these scam pathways. Payment-based scams also show substantial fear (35.7\%), aligning with immediate loss exposure, while ``Other'' scams exhibit lower fear (22.2\%) but comparatively higher surprise (33.3\%) and anger (22.2\%), which may reflect more abrupt deception cues or feelings of betrayal.

Notably, joy is uniquely observed in task-based scams (33.3\%) and is absent in payment and ``Other'' categories. While counterintuitive, this pattern may reflect early-stage reinforcement dynamics (e.g., small incentives or perceived progress) used to build trust and sustain engagement before harm is realized. Neutral responses appear in payment (21.4\%) and task-based scams (8.3\%), suggesting that a subset of narratives may be framed in a more detached or matter-of-fact tone rather than overt affective language.

Overall, the distributions suggest that scam structures may be associated with distinct emotional contours, reinforcing the need for prevention messaging and interventions that account for pathway-specific victim experiences. These results should be interpreted cautiously as exploratory descriptive evidence due to limited narrative sample size.

\subsection{Missing Data Diagnostics}

\subsubsection*{Missingness by Question Type}

Table~\ref{tab:missingness_appendix} reports average non-response rates by question category. The strong gradient supports the interpretation of non-response as a behavioral signal rather than random noise.\\

\begin{table}[!t]
\centering
\caption{Missingness Rates by Question Type}
\label{tab:missingness_appendix}
\small
\begin{tabular}{lc}
\hline
\textbf{Question Type} & \textbf{Missing (\%)} \\
\hline
Open-ended narratives & 95.5 \\
Financial loss amounts & 87.1 \\
Reporting behavior & 65.9 \\
Emotional responses & 63.2 \\
Demographics & 27.8 \\
\hline
\end{tabular}
\end{table}

\textbf{Conflict of Interest}: The authors declare that they have no known competing financial interests or personal relationships that could have appeared to influence the work reported in this paper.\\
\textbf{Data Availability}: The dataset and code are provided here in this anonymous GitHub: \\ \url{https://anonymous.4open.science/r/Job-scams-D3CD/README.md}.\\
\textbf{Funding}: This research received no specific grant from any funding agency, commercial, or not-for-profit sectors.\\
\textbf{AI-assisted copy editing}: The authors declare that AI-assisted copy editing was utilized in the preparation of this manuscript to enhance clarity and readability. However, all intellectual content and conclusions remain solely the responsibility of the authors.

\end{document}